\documentstyle[epsfig]{mn}

\title[Variability in the IMF]{Variability in the stellar initial mass
function at low and high mass: 3-component IMF models}

\author[Bruce G. Elmegreen]
{B. G. Elmegreen\thanks{E-mail: bge@watson.ibm.com}\\
IBM Research Division, T.J. Watson Research Center, P.O. Box 218,
Yorktown Heights, NY 10598, USA}

\date{Received  / Accepted  }

\begin{document}

\maketitle

\label{firstpage}

\markboth{Elmegreen: Variability in the IMF}{}

\begin{abstract}
Three component models of the IMF are made to consider possible
origins for the observed relative variations in the numbers of
brown dwarfs, solar-to-intermediate mass stars, and high mass
stars. The differences between the IMFs observed for clusters,
field, and remote field are also discussed. Three distinct
physical processes that should dominate the three stellar mass
regimes are noted.  The characteristic mass for most star
formation is identified with the thermal Jeans mass in the
molecular cloud core, and this presumably leads to the middle mass
range by the usual collapse and accretion processes. Pre-stellar
condensations (PSCs) observed in mm-wave continuum studies
presumably form at this mass. Significantly smaller
self-gravitating masses require much larger pressures and may
arise following dynamical processes inside these PSCs, including
disk formation, tight-cluster ejection, and photoevaporation as
studied elsewhere, but also gravitational collapse of shocked gas
in colliding PSCs. Significantly larger stellar masses form in
relatively low abundance by normal cloud processes, possibly
leading to steep IMFs in low-pressure field regions, but this
mass range can be significantly extended in high pressure cloud
cores by gravitationally-focussed gas accretion onto PSCs and by
the coalescence of PSCs.  
These models suggest that the observed variations in
brown dwarf, solar-to-intermediate mass, and high mass populations
are the result of dynamical effects that depend on environmental
density and velocity dispersion. 
They accommodate observations
ranging from shallow IMFs in cluster cores to Salpeter IMFs in
average clusters and whole galaxies to steep and even steeper IMFs
in field and remote field regions.  
They also suggest how the
top-heavy IMFs in some starburst clusters may originate and they
explain bottom-heavy IMFs in low surface brightness galaxies.
\end{abstract}

\begin{keywords}
stars: formation, stars: mass function, ISM: clouds
\end{keywords}

\section{Introduction}

The stellar initial mass function (IMF) is relatively uniform when
averaged over whole clusters or large regions of galaxies (see
review in Chabrier 2003), but detailed spatial variations in the
IMF that are greater than the statistical uncertainties suggest
there are different physical processes at work in at least three
mass regimes: brown dwarfs, solar to intermediate stellar masses,
and high mass stars.  Instead of a universal power law with a
lower mass limit, the IMF could be a composite with variable
contributions from at least these three regimes, depending on
environment. This paper reviews the observations of IMF variations
(Sect.\ref{sect:obs}), gives examples of how separate mass
functions may combine to produce the observed IMF (Sect.
\ref{sect:models}; see also Shadmehri 2004), 
and suggests how distinct physical processes
during star formation could lead to a 3-component IMF (Sect.
\ref{sect:phys}). 

\section{Observed Variations in the IMF}
\label{sect:obs}

\subsection{Low Mass Stars}
Several studies indicate that the relative abundance of brown
dwarfs and low mass stars compared to stars of intermediate mass
varies by a significant factor from region to region. For example,
IC 348 (Preibisch, Stanke \& Zinnecker 2003; Muench et al. 2003;
Luhman et al. 2003) and Taurus (Luhman 2000; Briceno et al. 2002)
have brown dwarf-to-star ratios that are $\sim2$ times lower than
the Orion trapezium cluster (Hillenbrand \& Carpenter 2000; Luhman
et al. 2000; Muench et al. 2002), Pleiades (Bouvier et al. 1998;
Luhman et al. 2000), M35 (Barrado y Navascu\'es et al. 2001), and
the galactic field (Reid et al. 1999). Even within the sub-solar
mass range, the IMF varies as a function of position within a
single cluster (IC 348: Muench et al. 2003; Orion: Hillenbrand
1997, Hillenbrand \& Carpenter 2000) and from cluster to cluster
(IC 348 versus Taurus; Luhman et al. 2003).

These low mass variations have been attributed to a dependence of
the Jeans mass on column density (Brice\~no et al. 2002) or Mach
number (Padoan \& Nordlund 2002), to stochasticity in the ages and
ejection rates of protostars from super-dense clusters (Reipurth
\& Clarke 2001; Bate, Bonnell \& Bromm 2002; Preibisch et al.
2003; Kroupa \& Bouvier 2003), to differences in the
photoevaporation rate from massive stellar neighbors (Preibisch et
al. 2003; Kroupa \& Bouvier 2003), and to possible differences in
the initial conditions of the turbulence (Delgado-Donate, Clarke,
\& Bate 2004). The low mass IMF can also be strongly affected by
variations in the binary fraction (Malkov \& Zinnecker 2001).

\subsection{High Mass Stars}
Two types of IMF variations have been reported for high mass
stars: a relatively high fraction compared to intermediate mass
stars in dense cluster cores, and a high fraction compared to
low-mass stars in some starbursts. In both cases the IMF shifts
toward more massive stars in higher-density regions. The shift
also continues toward lower densities, because field stars often
have steeper IMFs than clusters (Sect. \ref{sect:field}).

The flattening of the IMF in cluster cores is well documented.
Many studies also note that the youngest clusters have not yet had
time to segregate in mass by dynamical processes after the stars
form (Bonnell \& Davies 1998).  The massive stars appear to favor
the cluster cores from birth (Pandey, Mahra, \& Sagar 1992;
Subramaniam, Sagar, \& Bhatt 1993; Malumuth \& Heap 1994; Brandl
et al. 1996; Fischer et al. 1998; Hillenbrand \& Hartmann 1998;
Figer, McLean, \& Morris 1999; Le Duigou \& Kn\"odlseder 2002;
Stolte et al. 2002;  Sirianni et al. 2002; Muench et al. 2003;
Gouliermis et al. 2004; Lyo et al. 2004).  If $\Gamma$ is defined
to be $d\log n/d\log M$ for an IMF in equal-$\log$
intervals\footnote{In this paper, $\log$ refers to base 10},
$n(\log M)d\log M$, which gives $\Gamma=-1.35$ for the Salpeter
(1955) IMF, then cluster cores can have very flat IMFs,
$\Gamma\sim0$, and cluster envelopes can be much steeper,
$\Gamma\sim-2.5$ (de Grijs et al. 2002).

The observation of mass segregation at an earlier stage than
pre-main sequence stars would be interesting. The 850 $\mu$
sources in Ophiuchus (Johnstone et al. 2000) are mass-segregated,
for example (Elmegreen \& Krakowski 2001), even though they appear
to be too young to have moved a cluster radius. However, these are
much lower mass objects than OB stars and the statistical
uncertainties are large for the Johnstone et al. sample.  Thus,
the relevance of this result to high mass segregation in clusters
is unclear.

Starburst regions and clusters in other galaxies also have a
high-mass bias.  Either their IMFs are shallower than
$\Gamma=-1.35$ or they have a high value of the low-mass turnover,
such as several M$_\odot$ (Rieke et al. 1993). This turn-over is
typically $\sim0.5$ M$_\odot$ in the solar neighborhood (Kroupa
2001).  Modern observations of the starburst effect combine
dynamical cluster masses from stellar velocity dispersions and
cluster radii with luminosities and ages based on photometry. The
result is the light-to-mass ratio, and sometimes this is so high
that it can only be explained by a lack of low mass stars
(Sternberg 1998; Smith \& Gallagher 2001; Alonso-Herrero et al.
2001; McCrady, Gilbert \& Graham 2003; F\"orster Schreiber et al.
2003).

\subsection{Steep Field IMFs}
\label{sect:field}

At the other extreme, the IMF appears to steepen for field regions
away from massive dense clusters, although this observation is
subject to systematic errors. In the local field, $\Gamma\sim-1.7$
to $-1.8$ (Scalo 1986; Rana 1987; Kroupa, Tout, \& Gilmore 1993).
In the LMC field $\Gamma\sim-1.8\pm0.09$ (Parker et al. 1998) with
high statistical significance.  For the remote LMC and Milky Way
fields, $\Gamma\sim-4$ (Massey et al. 1995; Massey 2002). Hill et
al. (1994) found that the massive stars inside the Lucke \& Hodge
associations in the LMC have $\Gamma=-1.08\pm0.2$, whereas the
dispersed massive stars outside the associations have
$\Gamma=-1.74\pm0.3$.  Other comparisons between low and high
density regions in the LMC had the same result (Hill et al. 1995;
Parker et al. 1998). These studies are reminiscent of the paper by
Garmany et al. (1982), who found $\Gamma=-2.1$ outside the solar
circle and $\Gamma=-1.3$ inside, using a complete sample of stars
more massive than 20 M$_\odot$ within 2.5 kpc of the Sun. They
suggested the excess of massive stars inside the Solar circle came
from OB associations in the Carina and Cygnus spiral arms. A more
recent study by Casassus et al. (2000) using ultra-compact HII
regions throughout the galaxy finds little difference between the
IMF slopes inside and outside the solar circle, but they get the
steeper value, $\Gamma\sim-2$, everywhere.

Subsequent IMF models used this concept of a bimodal IMF to
explain other things too, such as the metallicity gradient and
star formation rate in the Galaxy (G\"usten \& Mezger 1983) and
the possibility that disk dark matter is composed of high-mass
stellar remnants (Larson 1986). Bi-modality is not so clearly
defined, however. For example, the giant association NGC 604 in
M33 has the Salpeter IMF (Gonz\'alez Delgado \& P\'erez 2000), as
do $\sim100$ star complexes in 20 galaxies studied by Sakhibov \&
Smirnov (2000).  So does the dense cluster R136 in the 30 Dor
region of the LMC as well as many other clusters (Massey \& Hunter
1998; Sagar, Munari, \& de Boer 2001), including h and $\chi$
Persei (Slesnick, Hillenbrand, \& Massey 2002), NGC 6611 (Belikov
et al. 2000),  NGC 1960 and NGC 2194 (Sanner et al. 2000).
However, the Upper Sco association has the steep IMF, with
$\Gamma=-1.6$ to $-1.8$ (Preibisch et al. 2002; see also Brown
1998).

Other possible observations of high-mass bias are more subtle.
There are apparently excess numbers of high mass stars in the
massive clusters NGC 3603 (N\"urnberger \& Petr-Gotzens 2002) and
W51 (Okumura et al. 2000) compared with extrapolations of the mass
functions at intermediate mass; i.e., the mass functions become
shallow at high mass. However, this result is very uncertain. For
NGC 3603, Sung \& Bessell (2004) found a relative depletion of
stars with $M>15$ M$_\odot$, rather than an excess. Scalo (1998)
suggested that the slope at high mass could be systematically
shallower than the slope at intermediate mass: $\Gamma\sim-1.7$
for 1--10 M$_\odot$ and $\Gamma \sim-1.3$ for 10--100 M$_\odot$.
Unfortunately, systematic uncertainties are often large in the
high-mass part of IMF (e.g., Scalo 1986; Massey 1998), and random
variations for small numbers of stars can reproduce the overall
scatter in most cluster IMFs (Elmegreen 1999a; Kroupa 2001).

Several peculiar observations on much smaller scales may be
showing the same density effects, although the statistical
uncertainties are much larger. For example, the dispersed embedded
stars in Orion seem to have a steeper IMF than the Trapezium
cluster (Ali \& Depoy 1995). Also, there is an apparent separation
between high and low mass pre-main sequence stars near SN 1987A in
the Large Magellanic Clouds (Panagia et al. 2000).

A recent study of the mass-to-light ratio in the disks of low
surface brightness galaxies suggests the entire IMF there is
steep, with $\Gamma=-2.85$ (Lee et al. 2004).  These galaxies are
likely to have very low pressures because of their low stellar and
gaseous column densities. Star formation in the low-density
``field'' mode could be pervasive.

\subsection{Origins of Steep Field IMFs}

Part of the difficulty with these field star measurements is that
even though the IMFs are observed to be relatively steep, they
could still originate from the same shallow IMFs found in dense
clusters after the clusters disperse.  This can happen in four
ways: special conditions for star formation in the field which
produce a physical dependence of the maximum stellar mass on the
cluster mass; differential drift of long-lived low-mass stars into
the field, differential evaporation of peripheral cluster stars
that have a steep IMF from birth, and inadequate corrections for
background stars.

Elmegreen (1999a) showed how a Salpeter IMF for clusters that form
in clouds with a power law distribution of masses could sum to
give a steeper IMF if massive clouds, which are statistically more
likely to form high mass stars, are destroyed more readily by
their stars than low mass clouds. One reason for this could be the
rapid increase in Lyman continuum luminosity with increasing
stellar mass, which actually works out numerically to give the
Massey et al. (1995) slope of $\Gamma\sim-4$ considering sampling
statistics and cloud binding energy. This differential destruction
was thought to apply to the field and not associations because the
ambient pressure in the field is often $\sim100$ times lower than
in associations, making the field clouds more susceptible to
internal disruption than the association clouds.  The model
applies even if the ages of the field stars are too young for them
to have drifted there from nearby associations (as observed by
Massey et al. 1995 and Parker et al. 2001).

Differential drift of the long-lived low-mass stars out of
dispersed clusters and into the field also steepens the field IMF,
provided the stars have enough time to do this (Elmegreen 1997,
1999a). Hoopes, Walterbos \& Bothun (2001) explained the steep
mass function required for diffuse interstellar ionization in
nearby galaxies this way. Tremonti et al. (2002) also found a
clear difference in the populations of stars in the field and
clusters of the dwarf starburst galaxy NGC 5253, with a Salpeter
IMF for the clusters and a steeper IMF for the field, and
explained this difference as a result of cluster dispersal into
the field after 10 My, when the most massive stars have
disappeared.

Differential evaporation of peripheral cluster stars is a natural
explanation for the steep IMF in the field because $\Gamma$ in the
outer parts of clusters is already low, $\Gamma\sim-3$ or lower
(de Grijs et al. 2002).  These are also the stars most likely to
fill the field because they are the most weakly bound to the
cluster. According to the figures in de Grijs et al., all stars
beyond $\sim20^{\prime\prime}$ in the LMC clusters NGC 1805 and
NGC 1818 have $\Gamma<-2$.  The question remains whether the stars
used for the field IMFs are too massive and short-lived to have
migrated there from dense clusters. Modelling of this process
seems to be required.

A fourth possibility for the steep field IMF in the LMC was
proposed by Parker et al. (2001), who noted that corrections for
field star contamination of cluster IMFs can reduce an apparent
$\Gamma$ from $-1.7$ to $-1.35$. Such corrections were not applied
to the field IMF, though.

\subsection{Superposition of IMFs}

Kroupa \& Weidner (2003) reconsidered the point in Elmegreen
(1999a) about superpositions of cluster IMFs and offered a
different conclusion. They suggested that superposition of
Salpeter IMFs with a cluster mass function slope of $-2.2$
explains the $\Gamma\sim-1.8$ field star IMF.  However, the
superposition of IMFs in a region gives the IMF of all of the
stars there, whether or not they are in clusters. If superposition
alone gives $\Gamma=-1.8$ for the field, then it should give the
same $-1.8$ for whole galaxies, including the clusters. Dispersal
of clusters into the field will not change the average IMF in a
galaxy. However, the summed IMFs for whole galaxies are not steep
like the field, they are usually Salpeter IMFs, so superposition
alone cannot explain the field IMF.

A review of observed galaxy IMFs was in Elmegreen (1999b). Recent
observations include the following: The stellar mass-to-light
ratio and the Tully-Fisher relation for galaxies were explained
with the Salpeter IMF and shown to be inconsistent with steeper
functions by Bell \& de Jong (2001). The colors in Blue Compact
Dwarf galaxies were also reproduced by the Salpeter IMF (Cair\'os
et al. 2002). The cosmic star formation history and metallicities
of galaxies were studied by Baldry \& Glazebrook (2003), who
concluded that only the Salpeter IMF was appropriate, ruling out
anything as steep at $\Gamma=-1.7$ for $M>0.5$ M$\odot$. Pipino \&
Matteucci (2004) also required the Salpeter IMF to explain the
photochemical evolution of elliptical galaxies. Similarly,
Rejkuba, Greggio \& Zoccali (2004) found the Salpeter IMF in the
halo of the galaxy NGC 5128 (Cen A).

The observation of a Salpeter IMF for whole galaxies and
essentially the same IMF or shallower for individual dense
clusters implies there is little connection between stellar mass
and cluster mass for most star formation.  If there were a
one-to-one correspondence between these masses, so that low mass
clusters formed only low mass stars while high mass clusters
formed both low and high mass stars, then the superposition of
cluster IMFs would be significantly steeper than the observed
galaxy IMFs. If only a small part of all star formation, such as
the loose clusters in the field, has a correspondence between
cluster mass and maximum stellar mass (as a result of specific
physical effects and not random sampling), then this
sub-population can have an IMF that is steeper than each cluster
IMF without affecting the galaxy-wide average.

\subsection{Considering a Real Difference between Field and
Cluster IMFs}

The IMF in low density clusters typical of the field could be
intrinsically steeper than the IMFs in dense clusters in OB
associations. It seems reasonable that the three main groups of
stellar mass -- brown dwarfs, solar-to-intermediate mass, and high
mass -- all form by different combinations of processes. Enhanced
gravitational focussing and rapid centralization can make high
mass stars grow by a much larger factor than low mass stars
through accretion of peripheral gas and coalescence of other
protostars in dense cluster cores. Accretion has been extensively
discussed by Zinnecker (1982), Larson (1999, 2002), Myers (2000),
Bonnell et al. (1997, 2001, 2004), Basu \& Jones (2004) and
others. Coalescence of other stars or protostars was discussed by
Zinnecker (1986), Larson (1990), Price \& Podsiadlowski (1995),
and Stahler, Palla, \& Ho (2000), Shadmehri (2004) and others. 
Coalescence after
accretion drag (Bonnell et al. 1998) or after accretion-induced
cloud core contraction (Bonnell, Bate, \& Zinnecker 1998; Bonnell
\& Bate 2002) also seem likely in view of recent simulations
showing these and other effects (Klessen 2001; Bate, Bonnell, \&
Bromm 2003; Bonnell, Bate \& Vine 2003; Gammie, et al. 2003; Li,
et al. 2004).

Accretion and coalescence make the IMF depend on environment. The
environment is also important for the confinement of stellar winds
and ionization during the collapse phase of massive-star formation
(Garay \& Lizano 1999; Yorke \& Sonnhalter 2002; Churchwell 2002;
McKee \& Tan 2003). These processes are enhanced in dense cloud
cores, so massive star formation could be biased to these regions.
This is one explanation for the flattening of the IMF in cluster
cores, along with dynamical effects (Giersz \& Heggie 1996;
Gerhard 2000; Kroupa, Aarseth \& Hurley 2001; Portegies-Zwart et
al. 2004). An important question is whether the average IMFs in
dense clusters are also flatter than the average IMFs in loose
stellar aggregates.

A critical observation would be a steep IMF in many combined
low-density regions like Taurus, which together contain enough
stars to sample the IMF out to the O-star range (see also
discussion in Luhman 2000). If the IMF has the Salpeter slope,
such sampling requires $50^{1.35}=200$ stars greater than 1
M$_\odot$ or $10^{1.35}=22$ stars greater than 5 $M_\odot$ to
include one star with a mass greater than 50 M$_\odot$. For only
22 stars between 5 and 50 M$_\odot$, however, the slope in the
derived IMF will be statistically uncertain by $\pm1.3$ (Elmegreen
1999a), which is too large to detect a difference between
$\Gamma=-1.35$ and $\Gamma=-1.7$. To make the IMF slope accurate
to $\pm0.35$ requires $\sim100$ stars in a mass interval spanning
a factor of 10 (Elmegreen 1999a). For this number of stars in the
5-50 M$_\odot$ interval, the corresponding total number of stars
and the total of the cluster masses would have to be $\sim4700$
stars and 5400 M$_\odot$ respectively for $\Gamma=-1.35$ down to
0.3 M$_\odot$. It follows that if a survey of newborn or embedded
stars far from dense clusters includes $\sim 4700$ stars down to
$\sim0.3$ M$_\odot$, then the difference between $\Gamma=-1.7$ and
$\Gamma=-1.35$ at high mass could be verified for such remote
regions.  We emphasize that the stars in such a survey have to be
very young to avoid contamination by differential drift of
low-mass association stars into the same fields. The velocity
dispersions of young stars in Taurus, for example, are probably
$>2$ km s$^{-1}$ (Frink et al. 1997), so field stars much older
than $\sim10$ My will begin to mix with their drifting
counterparts from OB associations, which probably have even larger
velocity dispersions.

Differences between dispersed and clustered protostars are already
evident. The mm-wave continuum sources in Taurus are extended like
classical isothermal spheres whereas the analogous sources in
Perseus are denser and more truncated like Bonner-Ebert spheres
(Motte \& Andr\'e 2001). Enhanced boundary pressures or source
interactions are evidently more important in the denser regions.

\section{Multi-component IMF models}
\label{sect:models}

We consider two examples where different parts of the IMF are
relatively independent. The point is to demonstrate that the
observed power law distributions ranging from solar mass to high
mass stars do not necessarily imply a single scale-free star
formation mechanism. The first model builds the entire IMF from
three log-normals, each with its own characteristic stellar mass,
and illustrates the observed variations in the IMF by varying
slightly the amplitudes or widths of these log-normals. The second
uses a log-normal function with density-dependent parameters to
match both an approximate log-normal IMF in the field and a
shallow power-law IMF for R136 in 30 Dor (Massey \& Hunter 1998).
A model in Shadmehri (2004) follows 
the coalescence of pre-stellar condensations
(PSC) in a cloud core, assuming these condensations form with a
log-normal mass function and collapse on their free-fall times.
Physical motivations for these models are discussed in section
\ref{sect:phys}.

The models have many input parameters so the observed IMFs can be
matched in different ways. The goal is not to fit the IMF with a
particular model, but to show that the IMF can be a composite of
IMFs from several different physical processes. These processes
should also have their own list of most-dominant physical
parameters and so, like our models, could also have a range of
possible outcomes.  Blended together or with poor sampling
statistics, they can produce what appears to be a universal IMF.
Only when viewed on small scales and short time intervals, and
when enough similar regions are included to have good sampling
statistics, will this physical diversity show up as a variable
IMF.  The observations mentioned above suggest this diversity is
beginning to appear.  Numerical simulations can check this further
by comparing the physical processes involved with the formation of
stars in the three mass ranges.

\subsection{Three log-normals}

\begin{figure}
\epsfig{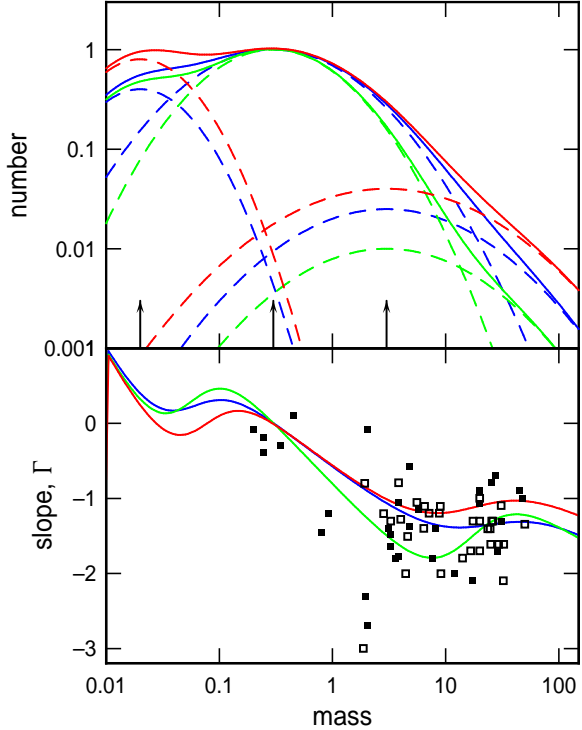} \caption{Three
component models of the IMF with distinct components indicated by
dashed lines. The top shows the IMFs and the bottom shows the
slopes along with observations from Scalo (1998) using filled
squares for the LMC and open squares for the Milky Way. Each
component is a log-normal with a characteristic amplitude, central
mass, and dispersion, as given in Table 1.}\label{fig:three}
\end{figure}

Figure \ref{fig:three}(top) shows examples of 3-part IMFs made
from the superposition of log-normal functions, one centered on
brown dwarfs at several hundredths of a solar mass, another at
several tenths of a solar mass, and a third at several solar
masses.  The bottom panel plots the slopes of the IMFs along with
open (Milky Way) and filled (LMC) squares that represent
observations (from Scalo 1998).

The composite IMF is based on the equation
\begin{eqnarray}
n(M)= A_1e^{-B_1\left(\log\left[M/M1\right]\right)^2}\nonumber\\
+A_2e^{-B_2\left(\log\left[M/M2\right]\right)^2}\nonumber\\
+A_3e^{-B_3\left(\log\left[M/M3\right]\right)^2}
\end{eqnarray}

Each of the components is indicated by a dashed line with
the same color as the composite. The model parameters are
summarized in Table 1.  The central masses for each
component are shown by arrows.

\begin{table*}
\centering
\caption{Model parameters for Figure 1}
\begin{tabular}{@{}lcccccccccccc@{}}
color & & $A_i$ & & & $M_i$ & & & $B_i$ & & & $\sigma_i$ & \\
& $i=1$ & 2 & 3 & $i=1$ & 2 & 3 & $i=1$ & 2 & 3 & $i=1$ & 2 & 3 \\
red   & 0.8 & 1.0 & 0.04 & 0.02 & 0.3 & 3.0 & 3.27 & 1.35 & 0.82 & 0.9 & 1.4 & 1.8 \\
blue  & 0.4 & 1.0 & 0.25 & 0.02 & 0.3 & 3.0 & 3.27 & 1.35 & 0.97 & 0.9 & 1.4 & 1.65 \\
green & 0.1 & 1.0 & 0.01 & 0.02 & 0.3 & 3.0 & 3.27 & 1.84 & 1.04 & 0.9 & 1.2 & 1.6
\end{tabular}
\end{table*}

For typical star counts in clusters, the observed IMFs would be
indistinguishable from these 3-component functions and probably
interpreted as power laws with a low mass turn-over. We choose
parameter values in the figure that give ``power-law'' slopes from
$\Gamma\sim-1.1$ to $-1.4$ at high mass and $-1.1$ to $-1.7$ at
intermediate mass. Observed variations over factors of 2 or 3 at
low mass (sect. I) have also been matched.  Thus this 3-component
model can reproduce the three-part variations that are observed in
the IMF.

\subsection{The upper IMF from a log-normal with density-dependent
dispersion}

\begin{figure}
\epsfig{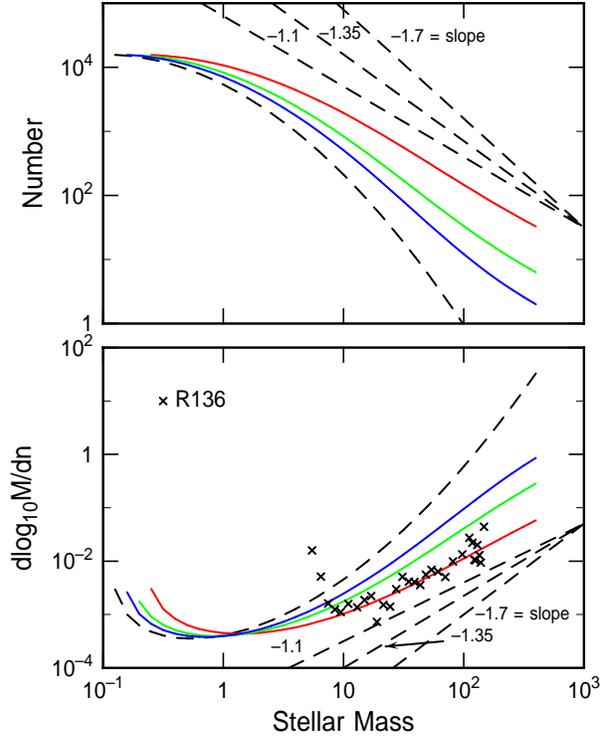} \caption{(top) IMF
model based on a log-normal mass distribution in which the
dispersion increases with density. The IMF is integrated over a
cloud density profile out to 2, 5, and 10 cloud core radii for
red, green, and blue colors (colors in electronic version only).  
(bottom) The mean separation between
the log of the masses in the IMF for the model shown in the top
and for the R136 cluster in the LMC.}\label{fig:buildd}
\end{figure}

Figure \ref{fig:buildd} shows a model where an idealized cloud
forms stars in a log-normal mass distribution that has a central
mass and dispersion increasing with cloud density.  The cloud
density has the form
\begin{equation}
\rho_c(r)=\left(1+\left[r/r_0\right]^2\right)^{-1}
\end{equation} where the central density is normalized to unity
and the core radius is $r_0$. The IMF is locally log-normal,
written as
\begin{equation}
f(M)=Ae^{-B\left(\log\left[M/M_0\right]\right)^2}\end{equation} for
exponential factor $B$ that decreases with density as
$B=B_1-B_2\rho_c(r)$ and central mass that increases with density
as $M_0=M_1+M_2\rho_c(r)$. Thus the log-normal is broader and
shifted toward higher mass in the cloud core.  The Miller-Scalo
(1979) IMF has $B_1=1.08$ and $M_1=0.1$ M$_\odot$ with no density
dependence. The parameters $B_2$ and $M_2$ are varied to fit the
observations. The overall stellar mass function in the cloud is
determined from an integral over cloud radius with a weighting
factor equal to the $3/2$ power of density:
\begin{equation}
n(M)={{\int_0^{R_{max}} f(M)\rho_c^{3/2} r^2dr}\over
{\int_0^{R_{max}} \rho_c^{3/2} r^2dr}}
\end{equation}
The density factor accounts for available mass and for a star
formation rate locally proportional to the dynamical rate,
$\left(G\rho_c\right)^{1/2}$.

Figure \ref{fig:buildd} shows the Miller-Scalo IMF ($B_2=M_2=0$)
as a dashed line and sample IMFs with $B_2=B_1$, $M_2=1,$ and
$R_{max}=2r_0$ (red), $5r_0$ (green), and $10r_0$ (blue). The IMF
slope is shallower for smaller $R_{max}$. The
bottom panel plots mean separation between the logs of the masses
in the IMF, or the inverse of the IMF density, which is defined to
be $d\log M/dn(M)$.  Also shown for comparison are the
observations of R136 from Massey \& Hunter (1998). This IMF
density is used to overcome problems with histograms or cumulative
distributions. When the number of stars is small, as for the R136
cluster at high mass, a histogram of the number of stars versus
log mass has gaps along the abscissa where certain stellar masses
happen to be missing, or it has such large mass intervals that the
slope of $n(M)$ has poor accuracy. To make Figure 2, we used the
599 stars with $M>5$ M$_\odot$ in the Massey \& Hunter survey
(kindly provided by the authors in electronic form). These stars
were listed in order of increasing mass and the differences in
$\log\left(M\right)$ for stellar groups of various lengths $N$
were determined. The IMF density is $\left(N-1\right)/\Delta
\log\left(M\right)$ and the inverse of this quantity is plotted.
When there are a large number of stars with a common mass, the
group size can be large.  The abscissa in this plot is the average
mass in the group. For example, if the two largest masses are
$M_a$ and $M_b$, then the last point plotted has coordinates
\begin{eqnarray}(x,y) & = & (\left[\log\left(M_b\right)
+\log\left(M_a\right)\right]/2,\nonumber \\\ & &\qquad\qquad
\left[\log\left(M_b\right)
-\log\left(M_a\right)\right]).\end{eqnarray} The slope of the
distribution of points in such a plot is the negative power
$-\Gamma$ in a power-law IMF.  The R136 points have
$\Gamma\sim-1.1$.

The variation of central mass and dispersion with density
stretches the IMF into a near-power law at high mass, with the
additional massive stars coming from the core region.  The slope
is more shallow as the core is approached, as shown by the
sequence of decreasing $R_{max}$ in the figure. The shallow IMF in
R136 can be reproduced in this way.

\section{Physical basis for multi-component models}
\label{sect:phys}

The stellar mass at the transition in the IMF where the rising
power law from high to intermediate mass bends over to become
relatively flat on a log-log plot has been viewed for some time as
the thermal Jeans mass in the cloud core, $M_{J0}$ (Larson 1992,
Elmegreen 1997, 1999a; see review in Chabrier 2003). This physical
mass is a reasonable lower limit to an IMF that arises from
scale-free turbulent fragmentation because fragmentation at lower
mass cannot easily lead to gravitational collapse for typical
cloud temperatures and pressures.  We believe that even without an
intrinsically scale-free star formation process, this
characteristic mass still defines the main component of the IMF
around $M_{J0}\sim0.1$ M$_\odot$. This component produces the
low-to-intermediate mass stars in all of the models above.

Stars or brown dwarfs with $0.1M_{J0}\sim0.01$ M$_\odot$ require
$100$ times higher pressures to collapse from a gaseous state than
do stars with $M_{J0}$ because the small stars need their own
thermal Jeans mass to be small for them to be made by an
instability, and the thermal Jeans mass scales as the inverse
square root of pressure for a fixed temperature. Several models of
brown dwarf formation were referenced in the introduction, i.e.
ejection from tight multiple systems or disks, photoevaporated
remains of normal protostars, and turbulent fragmentation. The
ejection and photoevaporation models do not rely on high {\it
cloud} pressure to reach a self-gravitating state with
$M<<M_{J0}$. They get their high pressures {\it inside}
pre-existing self-gravitating objects.

Ultra high pressures also occur during collisions between existing
gravitating pre-stellar condensations (PSC).  The PSCs with normal
stellar masses have high internal pressures from self-gravity but
no ability to fragment into independent (free-floating) brown
dwarfs. Whatever brown dwarfs they make will be in tight multiple
systems or disks, as in the existing models (these brown dwarfs
may be ejected after they form). However, two colliding PSCs will
have ultra-high pressure shocks between them, and the shock
fragments and sheared off overlapping parts can collapse into
independent brown dwarfs on their own. The free-fall time at the
compressed density is less than the cooling time by grain
radiation and probably also less than the timescale for fragment
dispersal. The pressure in such a collision is the square of the
collision speed times the PSC density, and because the average PSC
density is $\sim100$ times higher than the cloud core density
($10^7$ cm$^{-3}$ compared to $10^5$ cm$^{-3}$) while the
collision speed is probably comparable to the cloud core virial
speed, the pressure between colliding PSCs will be $\sim100$ times
the ambient cloud core pressure. This allows unstable pieces of
the collision to have a thermal Jeans mass that is 0.1 times the
thermal Jeans mass in the cloud core, making brown dwarfs.

Evidently, there are a variety of mechanisms, all related to
instabilities and fragments inside PSCs, that suggest a second
characteristic mass appears in the IMF at about 0.1 times the
characteristic mass in the average cloud core. The PSCs are
already dense and at high pressure because of self-gravity, and so
secondary fragmentation and collisions among them should make new
instabilities at $\sim10$ times lower $M_J$. This low
characteristic mass could be the origin of the low-mass part of
the 3-component IMF shown in Figure 1.

The high mass component of the IMF in this composite model may
also arise from collisions between PSCs if they occasionally
coalesce as well as fragment (Shadmehri 2004). 
Interactions could make both brown
dwarfs and massive PSCs at the same time if the ejected
gravitationally bound parts are a small fraction of the combined
mass. The above models assumed either a fixed characteristic mass
for the high mass part that was $\sim10$ times higher than the
main part of the IMF (Fig. \ref{fig:three}), or a variable
characteristic mass that depended on density (Fig.
\ref{fig:buildd}) or collision details (Shadmehri 2004). The
factor of $\sim10$ in mass is reasonable for distinct coalescence
bi-products because it follows from several collision events with
a hierarchical build-up in mass. Also, the gravitational cross
section for collisions becomes significantly larger than the
physical cross section, leading to a run-away growth process, when
one of the stars is $\sim10$ times the average mass of the others.

The exact factors for the high and low-mass components in our
composite models of the IMF are not critical because the
mathematical expressions are general enough to reproduce the
observed IMFs even with slightly different characteristic masses.
The essential points are: (1) there could be distinct physical
processes operating at low, intermediate, and high masses; (2)
these distinct processes will be disguised inside a composite IMF
if only wide spatial or temporal averages of stellar populations
are observed; and (3) the IMF need not reflect scale-free
formation conditions for all intermediate to massive stars.

\section{Summary}

Composite models of the IMF that are based on several
characteristic masses for distinct physical processes of star
formation are shown to be as good a fit to the observations as a
universal model that has essentially one characteristic mass and a
scale free formation process above that mass.  Three
characteristic masses and their possible origins were discussed
here: brown dwarf masses on the order of $0.02$ M$_\odot$ could be
the result of dynamical processes inside self-gravitating
pre-stellar condensations or gravitational collapse in the ultra
high-pressure shocks between these condensations;
small-to-intermediate mass stars could come from the pre-stellar
condensations themselves, getting their characteristic mass from
the thermal Jeans mass in the cloud core, and high mass stars
could grow from enhanced gas accretion and coalescence of
pre-stellar condensations (see also Shadmehri 2004). 
These three mass ranges would also have
processes in common, such as competitive accretion of ambient
cloud gas, collapse to protostars and disks, binary star
formation, and so on, as suggested by many authors referenced
above. These processes broaden each component in the IMF into what
was approximated here as a log-normal.

The characteristic mass for a star in this model is still the
thermal Jeans mass in the cloud core, $M_{J0}$.  This mass is
relatively free of variations over different environments in the
Galaxy and in different galaxies because of the way pressure and
temperature tend to scale for typical heating and cooling
conditions (Elmegreen 1999a). This constancy gives the apparent
universality of the IMF. However, secondary processes inside
$M_{J0}$ objects and interactions between these objects should
also form stars. With these additional star formation processes,
some of the observed variations in the IMF make more sense.  For
example, the relative number of brown dwarfs increases as the
primary condensations interact more and at greater relative speeds
in rich (Orion-type) clusters. Also, the relative numbers of
massive stars increase in denser, more coalescence-rich clusters,
including cluster cores and starbursts.

An important point here is that the shocks which form brown dwarfs
and the coalescences which form massive stars are among relatively
large ($10^3-10^4$ AU) pre-stellar condensations and not among
protostars or stars, which are much smaller. Thus the cluster
densities that lead to severe interactions are rather low, like
the observed densities of stars and protostars in clusters,
$10^3-10^4$ objects pc$^{-3}$ (Elmegreen \& Shadmehri 2003).

Computer simulations of cluster formation in turbulent clouds show
many processes operating simultaneously. It may be difficult to
tell if any particular process dominates during the formation of a
particular star. However, the IMFs in these simulations might
still be partitioned into the three fundamental parts discussed
here. Bate, Bonnell \& Bromm (2002) have already distinguished
brown dwarf formation from other stars. Gammie et al. (2003) have
shown how the high-mass part of the IMF gets shallower with time
as a result of coalescence and enhanced accretion. The advantage
of viewing the IMF in this 3-fold way is that it allows observers
to anticipate and recognize slight variations in the IMF for
different classes of regions when they are sampled with enough
stars to give statistically significant counts.

Acknowledgements: B.G.E. is grateful for support from NSF Grant
AST-0205097.

{}
\end{document}